\documentclass[12pt]{article}
\usepackage{amssymb}
\usepackage{epsfig}
\usepackage{graphicx}
\usepackage{amsmath}

\def \bea {\begin{eqnarray}}
\def \eea {\end{eqnarray}}
\def \bea* {\begin{eqnarray*}}
\def \eea* {\end{eqnarray*}}

\def \be {\begin{equation}}
\def \ee {\end{equation}}
\def \bes {\begin{equation*}}
\def \ees {\end{equation*}}

\def \pa  {\partial}
\def \f  {\frac}

\begin{document}

\begin{titlepage}

\begin{center}

\vskip 1.25 cm {\Large \bf The Schwinger Mechanism in (Anti) de Sitter Spacetimes}
\vskip 1.30 cm {P{\footnotesize RASANT} S{\footnotesize AMANTRAY}\footnote{prasant.samantray@iiti.ac.in}}\\
\vskip 0.50cm {\it Centre of Astronomy, Indian Institute of Technology Indore\\ Khandwa Road, Simrol 452020, India}

\end{center}

\vskip  .75 cm

\begin{abstract}
\baselineskip=16pt

\noindent We present a short and novel derivation of the Schwinger mechanism for particle pair production in $1+1$ dimensional de Sitter and Anti de Sitter spacetimes. We work directly in the flat embedding space and derive the pair production rates in these spacetimes via instanton methods. The derivation is manifestly coordinate independent, and also lends support to the possible deep connection between two conceptually disparate quantum phenomena - Schwinger effect and the Davies-Unruh effect. 

\end{abstract}

\end{titlepage}

\section{Introduction}

\noindent It is well known that in presence of an electric field, the vacuum is unstable and particle-antiparticle pair production occurs spontaneously \cite{Schwinger}. In flat Minkowski space, the probablity of this event occuring is given by 

\be
\Gamma \sim e^{\f{-\pi M^2}{eE}} \label{SchwingerEffect}
\ee  

\noindent where $e$ is the particle charge and $E$ is the constant external electric field. This probability is non-perturbative in the coupling ``$e$" as can be seen from (\ref{SchwingerEffect}). This effect, due to Schwinger, has been extensively explored in various contexts - especially in non-trivial backgrounds to understand the effects of temperature and/or spacetime curvature on the Schwinger mechanism \cite{Brown,Cai}. For instance, the Schwinger mechanism has been applied to particle production and false vacuum decay in de Sitter space \cite{Cai,Frob}. The de Sitter case is interesting from the cosmological perspective, and the computation essentially revolves around finding the one-loop effective action, formally via the heat kernel method. The Schwinger mechanism has also been investigated in Anti de Sitter backgrounds, in the context of charged Reissner-Nordstr\"{o}m (RN) black holes. RN black holes emit particles via Hawking radiation and since the near horizon geometry is $AdS_2 \times S^2$, there is an inevitable interplay between the Hawking process and Schwinger effect. \cite{Kim,Troost}. \\
\\
\noindent In addition to the Schwinger effect, quantum field theory predicts another remarkable and far reaching result - the Davies-Unruh effect \cite{Davies}. Stated simply, according to this effect an uniformly accelerating observer perceives a thermal bath with temperature proportional to its acceleration. Besides the fact that under the influence of a constant electric field, charged particles move with constant acceleration $a = \f{e E}{M}$, there does not seem to be any relation between the Schwinger and Davies-Unruh effects. However, the first hint of a relation between the two effects surfaces while studying Euclidean instantons in the context of Schwinger pair production. The instanton is characterized as a solution to the classical equations of motion, albeit in Euclidean time. Such a solution describes a closed circular orbit in flat Minkowski space. The corresponding action for an instanton in the presence of a constant electric field in flat space is given by $S_{Euc} = \f{\pi M^2}{e E}$, and the proper Euclidean time to complete this closed orbit is given by Hamilton-Jacobi relation $\tau_{Euc} = \pa_{M} S_{Euc} = \f{2\pi}{a} = \f{1}{T_{DU}}$, where $T_{DU}$ is the Davies-Unruh temperature. This points to a possible connection between Schwinger and the Davies-Unruh effect, and in this paper we advance this connection in both de Sitter and Anti de Sitter spacetimes. By working in the embedding space, we shall present a coordinate independent and unified treatment of deriving Schwinger effect in these spacetimes.\\
\\
\noindent We shall work in the semiclassical approximation using instanton methods. Additionally, we restrict ourselves to $1+1$ dimensions for simplicity. The qualitative picture for the Schwinger mechanism is as follows. Initially, there is just vacuum and the electric field is $E_{out}$ everywhere. Suddenly a particle-antiparticle pair is spontaneously created and the electric field drops to $E_{in}$ between the particles. Subsequently the pair accelerate apart, converting the electric field value to $E_{in}$ as they move away. The closed Euclidean worldline divides the space into ``inside" and ``outside" regions. On the worldline, the electric field is defined as the average sum of $E_{out}$ and $E_{in}$. In the instanton method, the charged particle couples to the electromagnetic field and by complexifying the time coordinate, the on-shell action (the coupling term and the surface term cancel each other on-shell) for the particle-field system is given by \cite{Garriga}

\be
S_{E} = M \int_{\Sigma} ds + \f{1}{4} \int_{Vol} F^{\mu\nu} F_{\mu\nu} \label{Eucl-action} 
\ee

\noindent We assume a constant external electric field. Following the work of Brown and Teitelboim \cite{Brown-Teitelboim}, we define the following quantities. 

\begin{eqnarray}
E_{out} - E_{in} &=& +e \nonumber \\
E_{out} + E_{in} &=& 2E_{on} \label{ElectricFields}
\end{eqnarray}

\noindent Using equations (\ref{Eucl-action}) and (\ref{ElectricFields}), the relevant instanton action is therefore given by 

\begin{eqnarray}
S_{E} [instanton] &=& S^{in}_{E} - S^{out}_{E} \nonumber \\
&=& M \int_{\Sigma} ds + \f{1}{2} (E^2_{in} - E^2_{out}) \int_{Vol} \nonumber \\
&=& M \int_{\Sigma} ds - e E_{on} \int_{Vol} \label{Instanton-action} 
\end{eqnarray}

\noindent Now consider a constant external electric field in dS$_2$/AdS$_2$ space given by
\begin{equation}
F_{\mu\nu} = -E_{out} \sqrt{-g} \epsilon_{\mu\nu} \label{ElectricField}
\end{equation}

\noindent where $\epsilon_{01} = - \epsilon_{10}=1$. Under the action of the above field, charged particles trace out worldlines according to the equations of motion
\be
a^{i}_{2} = \f{eF_{on}^{ij}u_{j}}{M} \label{EOM}
\ee

\noindent where $a^i_{2}$ is the 2-acceleration of the particle, $F_{on}^{ij}$ the field  strength defined on the worldline, and $u^i$ the usual 2-velocity. Using (\ref{ElectricField}) and (\ref{EOM}), and considering the fact that $u^i u_i = -1$, the magnitude of 2-acceleration is given by 

\be
a^{2}_{2} = g_{ij}u^i u^j = \frac{e^2 E_{on}^2}{M^2} \label{2-accel}
\ee

\noindent As defined before, $E_{on}$ is the electric field on the worldline.

\section{Schwinger Effect in dS$_2$}

\noindent Consider de Sitter space with scale $R$ defined by the hyperboloid

\be
-X^2_{0} + X^2_{1} + X^2_{2} = R^2  \label{dS-embedding}
\ee

\noindent From the perspective of embedding space, this hyperboloid exists in 2+1 dimensional flat Minkowski space. Consider a constant electric field along the direction $X_1$. The charged particle then accelerates along this direction keeping the coordinate $X_2 = X_c  = $ constant. Therefore, the trajectory in embedding space is given by

\be
-X^2_{0} + X^2_{1} = R^2 - X_{c}^2 \label{Trajectory-dS}
\ee 

\noindent suggesting that in embedding space the particle has an accelerating trajectory with ``3-acceleration" given by 

\be
a^{2}_{3} = 1/(R^2 - X_{c}^2) \label{3-accelDS}  
\ee 

\noindent In an elegant paper by Deser and Levin \cite{Deser-Levin}, it was shown that acceleration in embedding space is what determines the Davies-Unruh temperature in the target space as well. This equivalence of temperature is the key step. Therefore, in our case of dS$_2$, the Davies-Unruh temperature in terms of acceleration is given by

\begin{eqnarray}
a_3 &=& \sqrt{\f{1}{R^2} + a^{2}_{2}} \nonumber \\
&=& 2\pi T_{Unruh} \label{dSequiv} 
\end{eqnarray}

\noindent Therefore, from equations (\ref{2-accel}), (\ref{3-accelDS}) and (\ref{dSequiv}), we have 

\be
X_c = \f{e E_{on} R^2}{\sqrt{M^2 + e^2 E_{on}^2 R^2}} \label{dS-X}  
\ee

\noindent We can now compute the instanton action (\ref{Instanton-action}) in the embedding space. We first complexify the time coordinate as $X_0 \rightarrow i X_{0E}$, and thus from Eq.(\ref{Trajectory-dS}) the worldline radius becomes $R_0 = \sqrt{R^2 - X_{c}^2} = \f{M R}{\sqrt{M^2 + e^2 E_{on}^2 R^2}}$. The instanton action (\ref{Instanton-action}) for dS$_2$ can be calculated as 

\begin{eqnarray}
S_{E} [instanton] &=& 2 \pi M R_0 - e E_{on} \iiint_{D} \delta(\sqrt{X^2_{0E} + X^2_{1} + X^2_{2}} - R)~dX_{2} dX_{1} dX_{0E} \nonumber \\
&=& 2\pi R [\sqrt{M^2 + e^2 E_{on}^2 R^2} - e E_{on} R] \label{dS-instantonaction}  
\end{eqnarray}

\noindent where the domain of integration $D : -\sqrt{R^2 - X^2_{c}} \leq X_{0E} \leq \sqrt{R^2 - X^2_{c}};~~-\sqrt{R^2 - X^2_{c} - X^2_{0E}} \leq X_{1} \leq \sqrt{R^2 - X^2_{c} - X^2_{0E}};~~0 \leq X_{2} \leq \sqrt{R^2 - X^2_{1} - X^2_{0E}}$. Using (\ref{dS-instantonaction}), the Schwinger pair creation rate is given by

\be
\Gamma_{dS_2} \sim e^{ -2\pi R [\sqrt{M^2 + e^2 E_{on}^2 R^2} - e E_{on} R]} \label{dS-probability} 
\ee

\noindent We observe that even in the absence of any electric field, pair production occurs with rate $\Gamma_{dS_2} \sim e^{ -2\pi M R}$. This feature is the well known cosmological particle production of heavy fields in de Sitter space \cite{Cai,Frob,Garriga}. \\

\section{Schwinger Effect in AdS$_2$}

\noindent We now turn to Anti de Sitter space. Consider AdS$_2$ with scale $R$ defined by

\be
-X^2_{0} + X^2_{1} - X^2_{2} = -R^2  \label{AdS-embedding}
\ee

\noindent From the perspective of embedding space, this hyperboloid exists in 1+2 dimensional flat Minkowski space. This immediately raises a concern regarding closed time-like curves since both $X_0$ and $X_2$ behave like time coordinates. However, in our present work we sidestep this issue by taking $X_0$ as the ``relevant" time coordinate. We now consider a constant electric field along the direction $X_1$. The charged particle then accelerates along this direction keeping the coordinate $X_2 = X_c  = $ constant. Therefore, the trajectory is given by

\be
-X^2_{0} + X^2_{1} = X_{c}^2 - R^2 \label{Trajectory-AdS}
\ee 

\noindent This implies that in embedding space the particle has an accelerating trajectory with ``3-acceleration" given by 

\be
a^{2}_{3} = 1/(X_{c}^2 - R^2) \label{3-accelAdS}  
\ee 

\noindent Again following \cite{Deser-Levin}, the Davies-Unruh temperature in AdS$_2$ in terms of acceleration is given by

\begin{eqnarray}
a_3 &=& \sqrt{\f{-1}{R^2} + a^{2}_{2}} \nonumber \\
&=& 2\pi T_{Unruh} \label{AdSequiv} 
\end{eqnarray}

\noindent Therefore, from equations (\ref{2-accel}), (\ref{3-accelAdS}) and (\ref{AdSequiv}), we have 

\be
X_c = \f{e E_{on} R^2}{\sqrt{e^2 E_{on}^2 R^2 - M^2}} \label{AdS-X}  
\ee

\noindent The instanton action can now be computed as follows. As in the case of dS$_2$, the time coordinate is complexified as $X_0 \rightarrow i X_{0E}$, and thus from Eq.(\ref{Trajectory-AdS}) the worldline radius becomes $R_0 = \sqrt{X_{c}^2 - R^2} = \f{M R}{\sqrt{e^2 E_{on}^2 R^2 - M^2}}$. Therefore the instanton action (\ref{Instanton-action}) for AdS$_2$ can be calculated in the embedding space as 

\begin{equation}
S_{E} [instanton] = 2 \pi M R_0 - e E_{on} \iiint_{D} \delta(\sqrt{X^2_{2} - X^2_{1} - X^2_{0E}} - R)~dX_{2} dX_{1} dX_{0E}  
\end{equation}

\noindent where now the domain of integration is $D : -\sqrt{X^2_{c} - R^2} \leq X_{0E} \leq \sqrt{X^2_{c} - R^2};~~-\sqrt{X^2_{c} - X^2_{0E} - R^2} \leq X_{1} \leq \sqrt{X^2_{c} - X^2_{0E} - R^2};~~0 \leq X_{2} \leq \sqrt{X^2_{1} + X^2_{0E} + R^2}$. \\
\\
\noindent Evaluating the above integral near the neighborhood $\f{e E_{on} R}{\sqrt{e^2 E_{on}^2 R^2 - M^2}} \approx 1$, we get

\be
S_{E} [instanton] \approx 2\pi R [e E_{on} R - \sqrt{e^2 E_{on}^2 R^2 - M^2}] \label{AdS-instantonaction}
\ee

\noindent Therefore, the Schwinger pair creation rate is given by

\be
\Gamma_{AdS_2} \sim e^{ -2\pi R [e E_{on} R - \sqrt{e^2 E_{on}^2 R^2 - M^2}]} \label{AdS-probability} 
\ee

\noindent However, unlike in the case of dS$_2$, there exists a critical threshold electric field $E^2_{cr}  = \f{M^2}{e^2R^2}$ in AdS$_2$, below which there is no pair creation \cite{Kim,Troost}. This is due to the confining effects of AdS.

\section{Conclusion}

\noindent The derivation presented here is self-contained but a few concluding remarks are in order. In our calculation, we have ignored the gravitational backreaction, and have also implicitly assumed that mass of the instanton is much greater than curvature scales, i.e. $M^2 \gg \f{1}{R^2}$. If one relaxes this assumption, the instanton actions and pair production rates (\ref{dS-instantonaction}), (\ref{dS-probability}), (\ref{AdS-instantonaction}) and (\ref{AdS-probability}) will be modified by simply shifting the mass squared term to $M^2 \rightarrow M^2 \mp \f{1}{4R^2}$ for de Sitter and Anti de Sitter spaces respectively. This shift can be understood by looking at the quadratic Casimirs for de Sitter and Anti de Sitter groups, SO(1,2) and Sl(2,R), respectively.
\\
\noindent Though operationally we have used the kinematic equalities in Eqs.(\ref{dSequiv}) and (\ref{AdSequiv}) to derive our results, it is noteworthy to belabor on their interpretation. The first equality in both these equations is a classical relation, and in fact, the relation between accelerations in any embedding space and a submanifold of it ($a_D$ and $a_{D-1}$) is a purely differential geometric result (see \cite{Dawood} for a pedagogical discussion). However, the appearance of temperature, which has its origins in quantum statistics, presents a conceptually different interpretation of Eqs.(\ref{dSequiv}) and (\ref{AdSequiv}). There is no general theorem that establishes the equivalence of Unruh temperature for a general embedding space and its submanifold. Such an equivalence has only been established for (Anti) de Sitter spacetimes \cite{Deser-Levin}, where the detector response for target spaces as well as their corresponding embedding spaces was explicitly calculated using quantum field theory. We therefore posit that Eqs.(\ref{dSequiv}) and (\ref{AdSequiv}) also hold true at the quantum level, thereby latently encapsulating the Davies-Unruh effect in our derivation. \\

\noindent The connection with the Davies-Unruh effect also manifests itself in the following way. Pair creation rates in (A)dS$_2$ obtained previously by various authors used specific coordinate systems for simplification. Inflationary coordinates for de Sitter, and Poincare coordinates for Anti de Sitter spacetime were the usual choices. However, the exact pair creation rates being reproduced in our coordinate independent framework points to a unique vacuum choice for the Schwinger process in these spacetimes. To elucidate, consider dS$_2$ in Eq.(\ref{dS-embedding}) with the isometry group SO(1,2). In the embedding space, choosing a constant electric field along ``$X_1$" direction breaks the SO(1,2) isometry, and this points to a choice of vacuum different from the Minkowskian one annihilated by $i\frac{\partial}{\partial X_0}$. As noted from Eq.(\ref{Trajectory-dS}), the charged quanta follow hyperbolas in the $(X_0, X_1)$ embedding plane. Therefore, foliating our 2+1 dimensional flat Minkowskian space by (i) surfaces of constant $ -X^2_{0} + X^2_{1} + X^2_{2} = \gamma^2$, and (ii) surfaces of constant $-X^2_{0} + X^2_{1} = \xi^2>0$, gives

\be 
ds^2 = \frac{d\gamma^2}{1 - \frac{\xi^2}{\gamma^2}} - \frac{2\gamma \xi d\gamma d\xi}{\gamma^2 - \xi^2} + \left(-\kappa^2 \xi^2 dt^2 +  \frac{d\xi^2}{1 - \frac{\xi^2}{\gamma^2}}\right) \label{induced-dS}
\ee

\noindent where $\kappa$ is a constant introduced for dimensional reasons. The induced metric on $\gamma = R =$ constant surface is 

\be
ds^2_{ind} = -\kappa^2 \xi^2 dt^2 +  \frac{d\xi^2}{1 - \frac{\xi^2}{R^2}}
\ee   

\noindent which is just 1+1 dimensional de Sitter space expressed in static coordinates (describing an accelerating observer in global de Sitter space). Tangentially, had we foliated our flat space by surfaces of constant $-X^2_{0} + X^2_{1} = \xi^2<0$, we would have got Rindler-AdS$_2$  \cite{Parikh-Sam} as the induced metric, i.e. 

\be 
ds^2 = \frac{d\gamma^2}{1 + \frac{\xi^2}{\gamma^2}} + \frac{2\gamma \xi d\gamma d\xi}{\gamma^2 + \xi^2} + \left(-\kappa^2 \xi^2 dt^2 +  \frac{d\xi^2}{1 + \frac{\xi^2}{\gamma^2}}\right) \label{induced-AdS}
\ee

\noindent Therefore, the Klein-Gordon equation written in 2+1 dimensional flat metric (\ref{induced-dS}) picks $i\frac{\partial}{\partial t} = i \left( X_1\frac{\partial}{\partial X_0} + X_0\frac{\partial}{\partial X_1} \right)$ as the Hamiltonian which annihilates the vacuum. Coupled with the fact that acceleration in embedding space is in one-to-one correspondence with acceleration in the physical/target space, this is indeed the natural vacuum choice for quantization - the Davies-Unruh vacuum for de Sitter space. It can be shown that similar arguments go through for Anti de Sitter space as well, where the induced metric turns out to be 1+1 dimensional Rindler-AdS spacetime describing accelerating observers in global Anti de Sitter space. These ideas will be explored in future work \cite{Sup-Sam}.\\  

\noindent Summarizing, our present derivation of the Schwinger pair creation rate rests on two ingredients - 1) Equivalence of Euclidean action at the level of both embedding as well as target space. 2) The equivalence of temperature at the level of both embedding as well as target space. This strongly suggests that instead of working in target space, it should be possible to demonstrate Schwinger pair creation in AdS/dS by employing a quantum field theoretic approach in their flat embedding spaces. However, unlike in the case of de Sitter, one may run into technical difficulties while attempting a field theoretic derivation in Anti de Sitter spacetime. This is due to the presence of two time-like coordinates. Therefore, additional assumptions and boundary conditions must be specified for the derivation to go through \cite{Sup-Sam}. Additionally, while the Davies-Unruh effect is essentially thermodynamic in nature, the Schwinger effect is not. Though the possible connection between these two phenomena has been touched upon previously in the literature \cite{Cai,Kim,MVold,Parentani,Gavrilov,Juan}, we believe our present derivation of Schwinger pair creation mechanism to be the most concrete realization of this connection.
\section*{Acknowledgements} 
I thank Suprit Singh and Dawood Kothawala for useful discussions, and also thank IUCAA for its hospitality where this work was done.
   

\end{document}